\documentclass[aps,pre,onecolumn,showpacs,groupedaddress]{revtex4}

\usepackage{graphicx,psfrag}

\def\BibTeX{{\rm B\kern-.05em{\sc i\kern-.025em b}\kern-.08em
    T\kern-.1667em\lower.7ex\hbox{E}\kern-.125emX}}

\begin{document}

\title{Degree weighted recurrence networks for the analysis of time series data}

\author{RINKU JACOB}
\email{rinku.jacob.vallanat@gmail.com}
\affiliation{Department of Physics, The Cochin College, Cochin-682 002, India}
\author{K. P. HARIKRISHNAN}
\email{kp.hk05@gmail.com}
\affiliation{Department of Physics, The Cochin College, Cochin-682 002, India} 
\author{R. MISRA}
\email{rmisra@iucaa.in}
\affiliation{Inter University Centre for Astronomy and Astrophysics, Pune-411 007, India} 
\author{G. AMBIKA}
\email{g.ambika@iiserpune.ac.in}
\affiliation{Indian Institute of Science Education and Research, Pune-411 008, India} 

\begin{abstract}
Recurrence networks are powerful tools used effectively in the nonlinear analysis of time series data. The analysis in this context is done mostly with unweighted and undirected complex networks constructed with specific criteria from the time series. In this work, we propose a novel method to construct ``weighted recurrence network''(WRN) from a time series and show how it can reveal useful information regarding the structure of a chaotic attractor, which the usual unweighted recurrence network cannot provide. Especially, we find the node strength distribution of the WRN, from every chaotic attractor follows a power law (with exponential tail) with the index characteristic to the fractal structure of the attractor. This leads to a new class among complex networks, to which networks from all standard chaotic attractors are found to belong. In addition, we present generalized definitions for clustering coefficient and characteristic path length and show that these measures can effectively discriminate chaotic dynamics from white noise and $1/f$ colored noise. Our results indicate that the WRN and the associated measures can become potentially important tools for the analysis of short and noisy time series from the real world systems as they are clearly demarked from that of noisy or stochastic systems.     
\end{abstract}


\maketitle

\section{\label{sec:level1}INTRODUCTION}
Analysis of time series data using complex network measures has become an important area of research over the last two decades \cite {alb}. 
Several methods \cite {zha,xu,mar} have been proposed in the literature to transform a time series data into a complex network, with each of them finding application in particular contexts. The measures derived from the resultant recurrence networks provide information about the nature of dynamics underlying the time series. Moreover, this analysis is especially useful in real world applications with very small data or limited number of observations where the conventional approach using correlation dimension and entropy may give unreliable results.

A simple and direct method to convert time series to complex network is using the property of recurrence \cite {eck} of every dynamical system and the resulting network is called recurrence network (RN) \cite {don1}. In this, the time series is first embedded in a multi-variate state space of dimension $M$ using the time delay co-ordinates \cite {gra}. With every point in the attractor  as a node, a recurrence threshold ($\epsilon$) is set to define the connection between two nodes, if they fall within the threshold. This RN is an unweighted and undirected network with the elements of the adjacency matrix $A_{\imath \jmath}$ either $1$ or $0$ depending on whether two nodes are connected or not. Once constructed, the statistical measures 
\cite {don2} of the network characterize the structural properties of the attractor underlying the time series \cite {don3}.

The analysis of time series data reported so far using such RNs have been confined to unweighted networks. Very recently, Sun et al. \cite {sun} have stressed the importance of weighted networks in the analysis of time series from dynamical systems. They use a sliding window and symbolic scheme to construct the weighted network from time series.

In this paper, we present a novel method to construct weighted 
recurrence networks (WRN)from a time series. This construction for the first time could identify the WRNs from different chaotic attractors as a single class 
with similar distribution which may open up a new window in the study of complexity of dynamical systems through time series. The measures derived from WRNs of chaotic systems are clearly demarked from that of noisy or stochastic systems.
Our paper is organized as follows: The details regarding the 
the scheme for constructing the WRN are presented in the next section.  
In \S III, important network measures derived from the WRN and their relevance in characterizing the 
structural properties of chaotic attractors are discussed. The paper is concluded in \S IV.

\section{\label{sec:level1}WEIGHTED RECURRENCE NETWORKS FROM TIME SERIES}
\subsection{Selection of recurrence threshold}
The recurrence threshold, $\epsilon$, is a crucial parameter since the characteristic properties of the RN 
depend on its value. In general, for each embedded attractor from the time series, the value of 
$\epsilon$ has to be determined separately as it varies with the size of the attractor. Two 
criteria are usually employed \cite {don1,don2} to select $\epsilon$. The first and the primary one 
is that there should be a giant component for the resulting RN which sets a lower bound for 
$\epsilon$. In order to ensure that the network is not overconnected, the upper bound for $\epsilon$ 
is set such that the link density (the ratio of actual connections to all possible connections in a 
network of $N$ nodes) is only a small fixed fraction of the maximum possible value. This provides a 
small range $\Delta \epsilon$ of optimum threshold for each system where the resulting network is 
considered to be a proper network representation of the time series.

Recently, we have proposed a scheme \cite {rj1} where we tried to fix a small uniform range 
$\Delta \epsilon$ for choosing the threshold for  time series from different chaotic systems. For 
this, we first transform the time series to a uniform deviate so that the size of the attractor 
always remains within the unit cube. To find the lower bound of $\epsilon$, we use the standand 
criterion that the network turns into a single giant component. The upper bound is determined by 
the condition that the network is not overconnected. However, instead of fixing the link density, 
we apply a criterion that the characteristic path length (that defines the global connectivity of 
the network) of RN from chaotic time series is significantly different from that of white noise. 
This is because, the recurrence threshold is a property of dynamical systems only and hence a 
proper value of the threshold should make the network measures for RN from chaotic systems different 
from that of purely random data. Though this condition appears subjective, we are able to fix an 
empirical upper bound for $\Delta \epsilon$ with this. More importantly, with these 
criteria, it is possible to provide an approximately identical range of $\Delta \epsilon$ for 
constructing RN from different time series for a given embedding dimension $M$ if the number of 
nodes $N$ in the network is $< 10000$. 

It should be noted that the main motivation behind the above scheme for the choice of recurrence 
threshold is to search for a uniform framework for recurrence network analysis so that comparison of 
network measures between different systems is possible. Though the exact value of $\epsilon$ at 
which the RN just becomes a giant component slightly differs for different systems, we are able to 
achieve a certain level of automation in the analysis as a reward for this small compromise. The 
scheme has been effectively applied to compare network measures from different chaotic attractors 
\cite {rj1}, to study the influence of noise on the structure of chaotic attractors \cite {rj2} and 
to propose a new heterogeneity index \cite {rj3} for complex networks which, in turn, provides a 
unique measure for each chaotic attrator through RN. We stick to the same criteria for the 
selection of $\epsilon$ in this work. 

\subsection{Construction of the weighted recurrence network}
The unweighted RN is constructed first from the time series after embedding using the selected value of $\epsilon$. In order to convert 
it into a weighted RN, one has to assign weight factor to every link in the network. For weighted 
networks that model any real world system or interaction, the weight factor will be specific to the 
network. For example, in a transportation network, it may depend on the distance between two 
nodes while for a communication network, the same may be characterized by the rate of information 
transfer through the link. Here we introduce a general criterion, based on the intrinsic topology itself, for assigning the weight factors and hence can be adopted to any kind of network. We show how this method is especially befitting for RNs and its related complexity measures.

Starting with a RN of $N$ number of nodes and the $\imath^{th}$ node has a degree $k_i$ the weight $w_{ij}$ for the link 
between two nodes $\imath$ and $\jmath$ in the network is defined as:
\begin{equation}
w_{ij} = {{\sqrt {k_i k_j}} \over {k_{max}}}
\label{eq:1}
\end{equation}
where $k_{max}$ is the maximum degree in the network. Note that the maximum possible value of 
$w_{ij}$ is normalized as $1$ and occurs for a link between two nodes which are connected to 
$k_{max}$ other nodes in the network. For a reference node $i$ in general, it is connected to 
$k_i$ other nodes with each link having a different weight factor. Similar weight distributions 
have been shown to exist in certain real world networks as well \cite {barr}. 

The above method of assigning the weight factor is a non-subjective 
criterion (independent of the details of interaction the network represents) and hence can be 
adopted in several contexts. In particular, in the case of RNs, while the connection between two nodes comes by way of proximity of the corresponding points on the embedded attractor, in WRN, the weights also pick up the extent of clustering around each point. Hence WRN can be a better tool to characterize the probability density variations over the attractor, as is clear from the study presented here.

The sum of the weight factors associated with a node as determined by its connections is 
defined as the \emph {strength} of the node, $s$ \cite {new1,ops1}.  
For example, for the node $\imath$, we have: 
\begin{equation}
s(i) = \sum_{j=1}^{k_i} w_{ij}
\label{eq:2}
\end{equation} 
If all the nodes have equal number of connections $<k>$, the weight factor of each node is 
approximately the same and the network can be considered as a homogeneous weighted network. 
As the weight factors among the nodes become more diverse, the network becomes more 
heterogeneous. The average weight factor associated with the whole network is defined as the 
weighted link density:
\begin{equation}
\rho_w = {{\sum_{i,j} w_{ij}} \over {N(N-1)}}
\label{eq:3}
\end{equation}

\section{\label{sec:level1}MEASURES FROM WEIGHTED RECURRENCE NETWORK} 
After constructing the WRN, we are now in a position to analyze various time series data using the 
characteristic measures of WRN. Here we generalize three important network measures, the degree 
distribution, the clustering coefficient (CC) and the characteristic path length (CPL) for the WRN. 

For our analysis, we use time series from two standard chaotic systems and two representative noise 
data as examples to illustrate the potential of WRN and its utility in the analysis of time series 
data. The chaotic time series are from the standard Lorenz attractor (parameters $\sigma = 10$, 
$\rho = 8/3$ and $r = 28$) and the R\"ossler attractor (parameters $a = 0.2$, $b = 0.2$ 
and $c = 7.8$). In both cases, the time series is generated using a time step $\Delta t = 0.05$. 

Two different types of noise data are also used for the analysis. One is the white noise and the 
other one is a candidate from the family of colored noise. The colored noise are correlated 
random processes which are considered to be important in the analysis of chaotic data since they 
share many characteristic properties common to chaotic data \cite {osb}. 
They are a class of random 
processes with power law varying as $1/f^{\alpha}$, with $\alpha$ ranging, typically, from $1$ to $2$. 
Here we choose $1/f$ noise as a representative colored noise which are ubiquitous in the real 
world and are popularly called brownian noise, since they behave identical to the brownian motion. 
To compute the characteristic measures, $10$ different time series are used (by changing initial 
conditions for chaos and with different simulations for noise) and the average is taken.

\begin{figure}
\includegraphics[width=0.90\columnwidth]{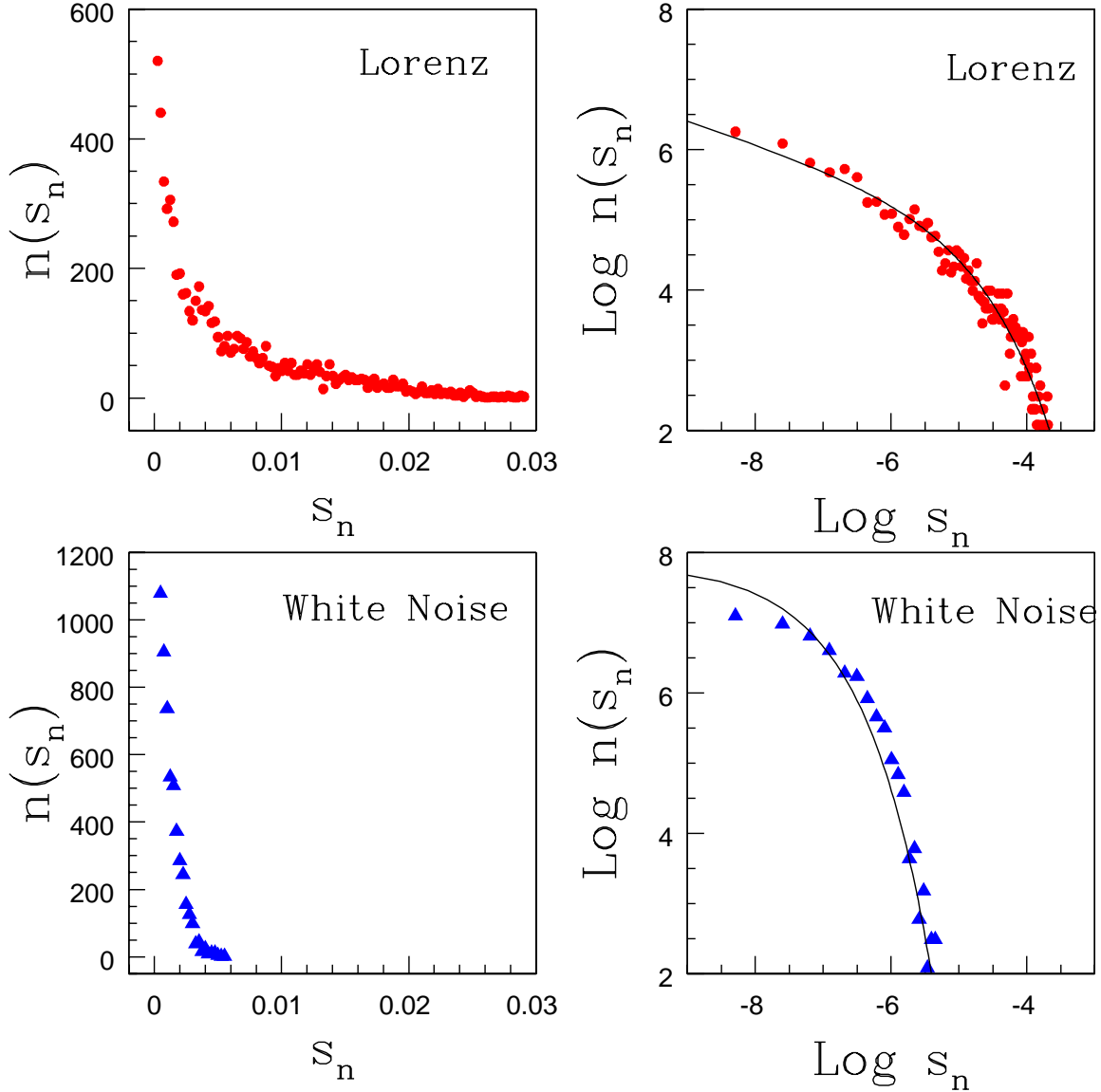}%
\caption{\label{f.1}(Color online) Normalized node strength distribution in the weighted RN constructed 
from the Lorenz attractor time series (red solid circles) and white noise (blue solid triangles) are 
shown in the left panel. The same distributions drawn in the log scale are shown in the right panel. The 
solid line in the right panel represents the fitting function for the corresponding distributions (see text). 
The embedding dimension used for constructing the networks is $M = 3$ and the number of nodes in the network, 
$N = 5000$.}
\label{f.1}
\end{figure}
  
\subsection{Normalized strength distribution}
For any unweighted complex network, the degree distribution, denoted by $P(k)$, is a probability 
distribution representing how many nodes have a given degree $k$. For random graphs (RG), the degree 
distribution is Poissonian where as for scale free (SF) networks, it obeys a power law \cite {new}. 
For the RN from chaotic time series, the degree distribution is characteristic to the structure of 
the attractor \cite {rj1}. To generalize the degree distribution for the WRN, we first note that the 
characteristic property of a node that decides its connectivity in the network is not its degree, but 
its strength $s$ as defined in  Eq.~\ref{eq:2}. In other words, the degree distribution has to be 
replaced by the \emph {strength distribution} \cite {noh} of the weighted network which represents the 
probability $P(s)$ of nodes having a given strength $s$ in a network of $N$ nodes. Even though $s$ varies 
discretely, it is not an integer like the degree $k$. 

Instead of using the strength distribution directly, we compute a \emph {normalized strength distribution} 
that reveals the utility of WRN. Since $s$ varies discretely, we can write 
\begin{equation}
\sum_s P(s) \Delta s = 1
\label{eq:4}
\end{equation}
We now find the number of nodes $n(s_n) \equiv N P(s)$ (rather than the probability of nodes) 
corresponding to a normalized strength $s_n = {{\Delta s} \over N}$ and the above equation can be 
re-written as 
\begin{equation}
\sum_{s_n} n(s_n) s_n = 1
\label{eq:5}
\end{equation}
Here $n(s_n)$ is the number of nodes having strength around the normalized value $s_n$, which varies 
in the range $[0,1]$. We now compute the normalized strength distribution for WRN from chaotic and 
random time series.

First we show that the distribution is qualitatively different for chaotic and random data. 
In Fig.~\ref{f.1} (left panel), we show the normalized node strength distribution of the WRN from the Lorenz 
attractor (solid circles) time series and the same for white noise (solid triangles). We have 
found that in the case of Lorenz, $n(s_n)$ decreases with $s_n$ as a power law initially with an 
exponantial cut off at the tail and the variation can be represented 
using the following functional fit:
\begin{equation}
n(s_n) \propto s_n^{-\gamma} e^{-s_n/c}
\label{eq:6}
\end{equation}
with the parameters $\gamma$ and $c$ depending on the particular system. 
On the other hand, the distribution for the WRN from white noise is found to be qualitatively 
different from that of chaotic time series. Here the power law part is absent and the function fit 
is purely exponential as: 
\begin{equation}
n(s_n) \propto  e^{-s_n/c}
\label{eq:7}
\end{equation}
For clarity, the same distributions are shown in log scale in Fig.~\ref{f.1} (right panel) with the 
functional fit as solid line in both cases.
 
In Fig.~\ref{f.2}, the normalized node strength distribution of the WRN from the Lorenz 
attractor (solid circles) and the R\"ossler attractor (solid triangles) are shown in  $\log$ scale  
along with the functional fit (solid line) in both cases. Similar result is obtained for many other 
standard chaotic attractors as well. The crucial parameter here is the power law index 
$\gamma$ indicating a scale free character for the distribution initially. The average value of $\gamma$ 
from $10$ different simulations for WRN from a few standard chaotic attractors are shown in Table I.

\begin{table}[h]
\centering
\begin{tabular}{|l|c|}
\hline
\emph{System} & $\gamma$ \\
\hline
\hline

Lorenz & 0.33 $\pm$ 0.05 \\

&  \\

R\"ossler & 0.14 $\pm$ 0.04 \\

&  \\

Ueda & 0.26 $\pm$ 0.05 \\

&  \\

Henon & 0.18 $\pm$ 0.03 \\

&  \\

Lozi & 0.20 $\pm$ 0.03 \\

\hline
\hline
\end{tabular}
\caption{Values of power law exponent in the node strength distribution for WRN from 
standard chaotic attractors.}
\label{tab:1}
\end{table}

We now show that this distribution is a characteristic property of every chaotic attractor and is 
independent of changes in parameters, such as, embedding dimension $M$ and the number of nodes in 
the network $N$. This is illustrated using Lorenz and R\"ossler attractors in Fig.~\ref{f.3}   
for two $N$ values with fixed $M$ and vice versa. The result  
implies that the power law index $\gamma$ is a characteristic index for a chaotic attractor. 

\begin{figure}
\includegraphics[width=0.90\columnwidth]{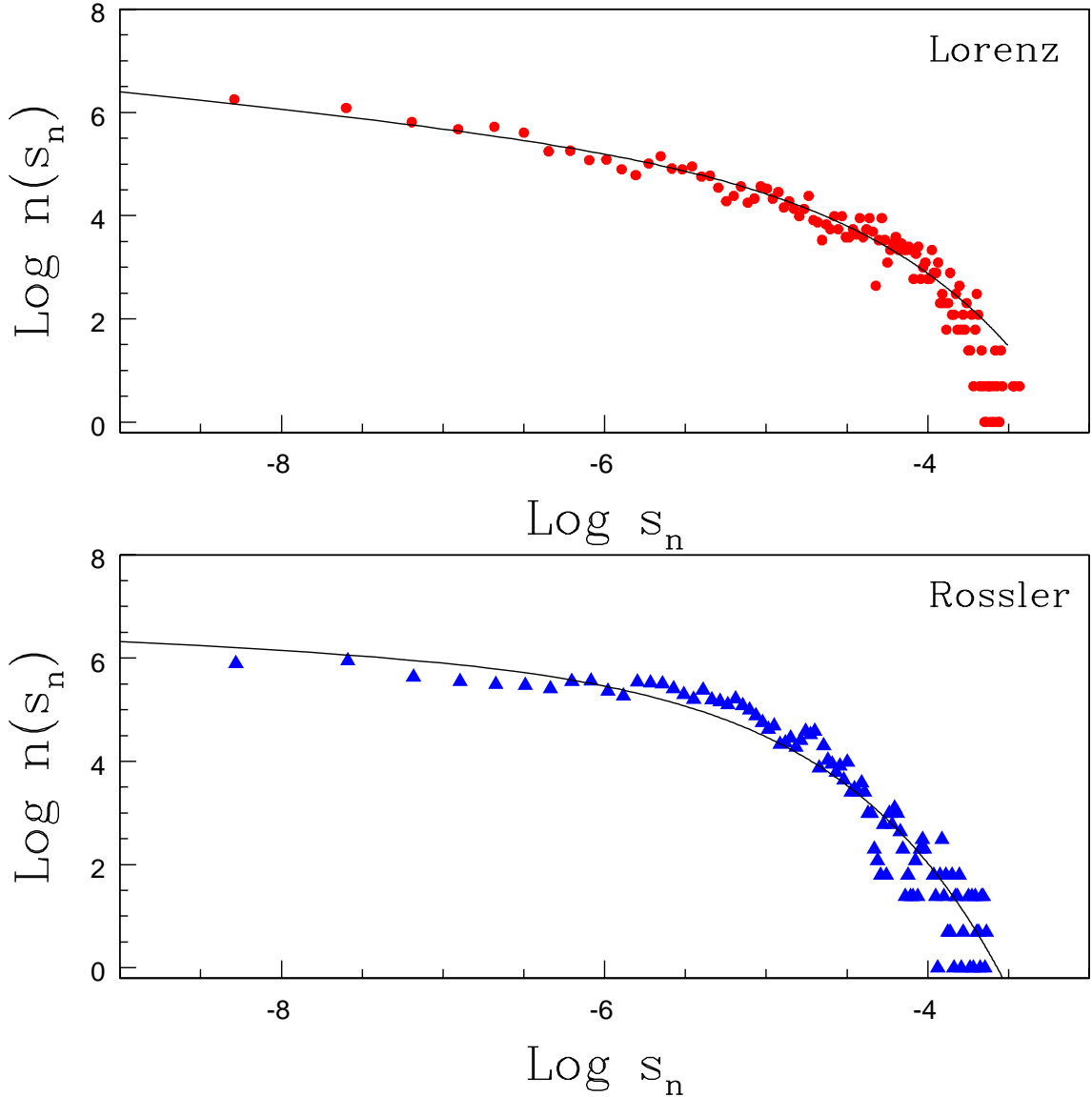}%
\caption{\label{f.2}(Color online)Normalized node strength distribution of the WRN constructed 
from the Lorenz attractor time series (red solid circles) in log scale is shown in the top panel. The 
solid line is the functional fit. The same for the R\"ossler 
attractor is shown in the bottom panel. The embedding dimension used for constructing the networks is 
$M = 3$ and the number of nodes in the network, $N = 5000$ in both cases.}
\label{f.2}
\end{figure}

\begin{figure}
\includegraphics[width=0.90\columnwidth]{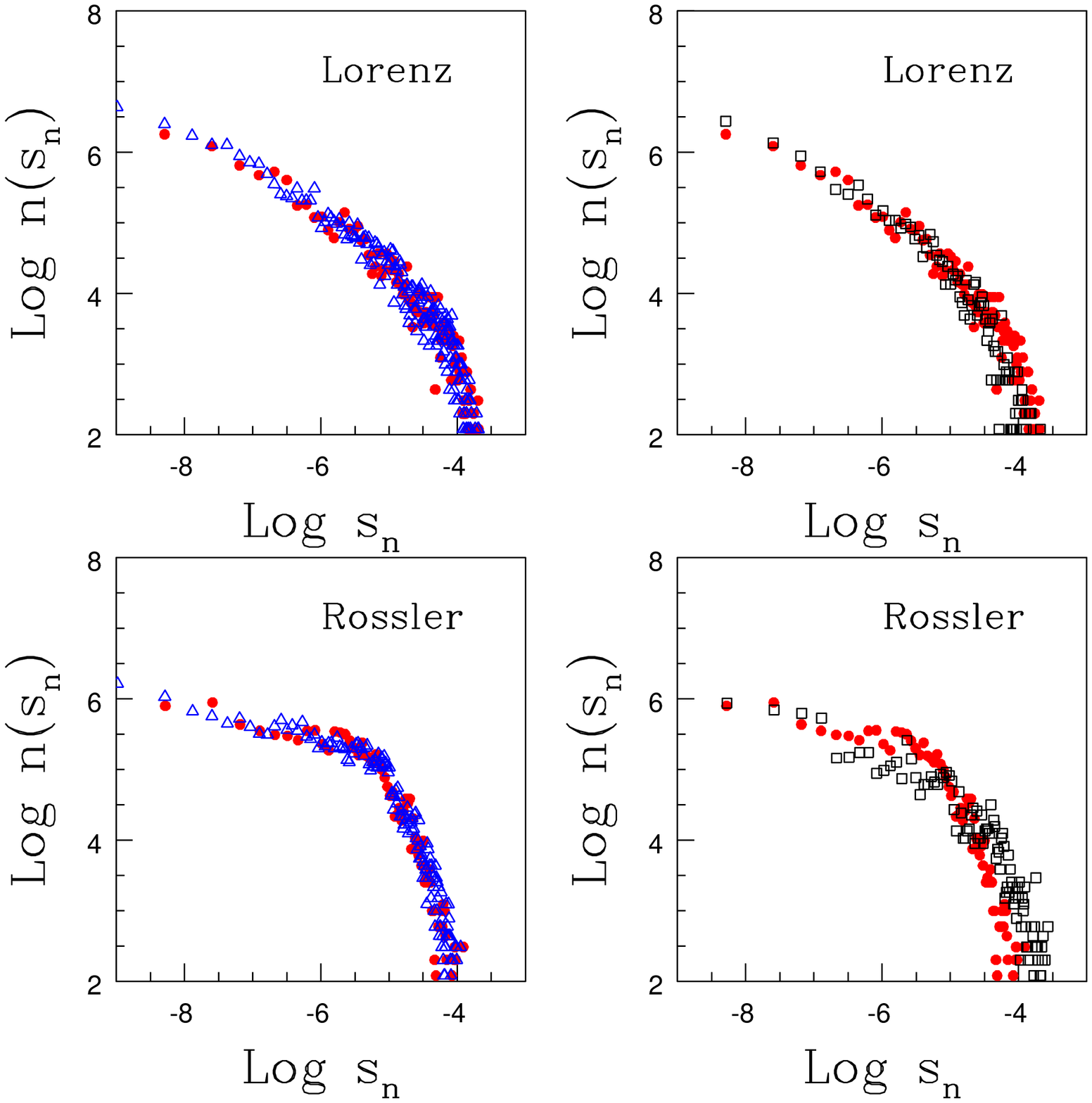}%
\caption{\label{f.3}(Color online)Node strength distribution of WRN constructed from two standard chaotic 
attractors for two different values of $N$ with fixed $M$ (left panel) and vice versa (right panel).  It is 
clear that the distribution remains unchanged with respect to changes in $M$ and $N$. The left panel is for 
$M = 3$ with $N = 5000$ (red solid circles) and $N = 10000$ (blue open triangles) while the right panel is 
for $N = 5000$ with $M = 3$ (red solid circles) and $M = 4$ (black open squares).}
\label{f.3}
\end{figure}

The computation is repeated for white noise and $1/f$ noise by changing $N$ and $M$ for both cases.  
In Fig.~\ref{f.4}, we show the result for white noise 
and $1/f$ noise for two $N$ values with fixed $M$ and vice versa.  Apart from the absence of power 
law scaling, the distribution is found to shift (lower panel) in both cases as $M$ changes. 
This is because, unlike the chaotic attractor, noise tends to fill the available state space which, 
in turn, changes the constructed network and the corresponding measures as $M$ changes.  

To understand the emergence of power law in the strength distribution for chaotic systems, we have 
looked at the construction of the WRN more closely and found that the scale free character follows 
from the method of finding the strength of a node by adding the weight factors of its links.  
For the unweighted RN, the degree $k_i$ of $\imath^{th}$ node represents the local probability 
density around the corresponding point on the attractor. So the degree distribution $P(k)$ 
approximately represents, as a discrete distribution, the variation in the number of (local) 
regions with a given probability density. This will be characteristic of the structure of the 
chaotic attractor and does not change with either $N$ or $M$, as we have shown \cite {rj1}. 

\begin{figure}
\includegraphics[width=0.90\columnwidth]{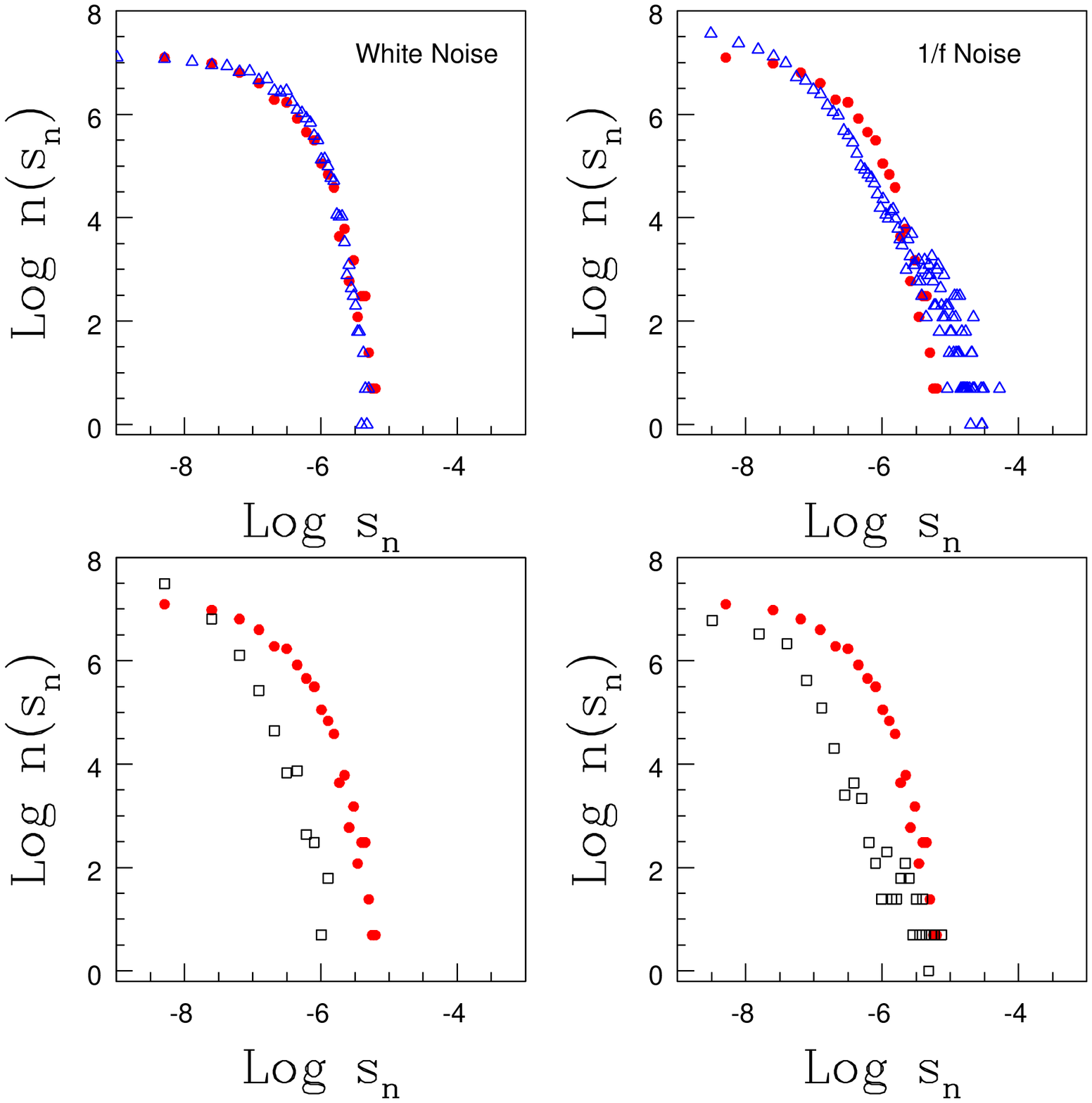}%
\caption{\label{f.4}(Color online) Node strength distribution of the WRN from white noise 
in the left panel (top and bottom) and $1/f$ noise in the right panel.  
In both cases, top panel shows the distribution for $N = 5000$ (red solid circles) 
and $N = 10000$ (blue open triangles) for $M$ fixed at $3$. Bottom panel shows the same for 
$M = 3$ (red solid circles) and $M = 4$ (black open squares) for $N = 5000$. Note the shift in the 
distribution in the bottom panel for both cases as $M$ changes.}
\label{f.4}
\end{figure}

\begin{figure}
\includegraphics[width=0.90\columnwidth]{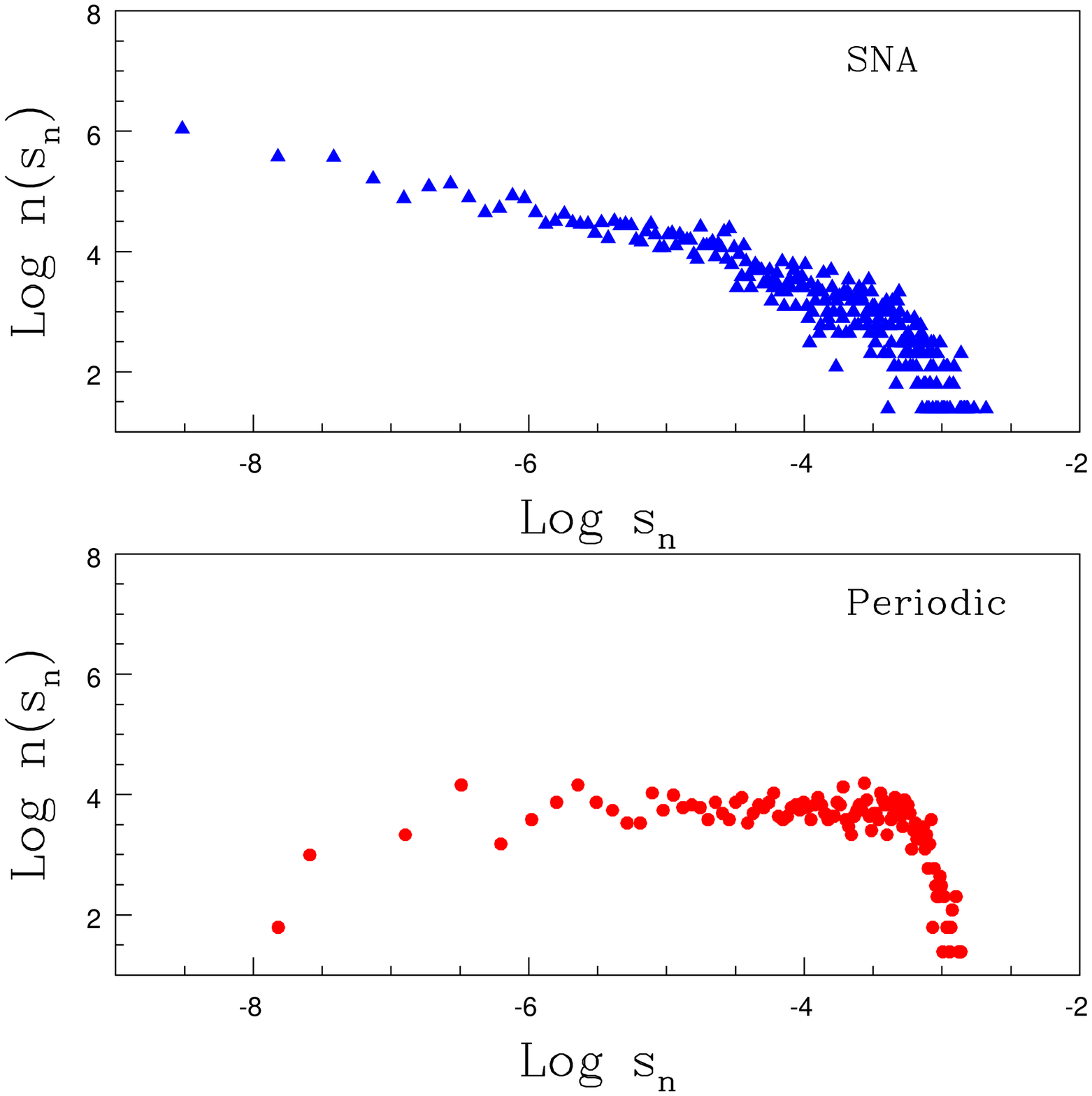}%
\caption{\label{f.5}(Color online) Normalized strength distribution of the WRN constructed from 
the time series of a strange nonchaotic attractor (see text) and that from a periodic attractor 
(bottom panel) from the R\"ossler system with $a = 0.1$, $b = 0.1$ and $c = 3.5$.}
\label{f.5}
\end{figure}

On the other hand, for the WRN with weight given by Eq.~\ref{eq:1}, each node has a strength $s$, 
which is the sum of the normalized weight factors of its connections. In other words, it represents 
a blunt measure as it takes into consideration a node's total involvement in the network and not the 
number of other nodes connected to it \cite {ops1}. For example, two nodes having totally different 
number of connections (that is, degree) can have the same node strength as the weight of edges of 
each connection are different. Thus, unlike the degree distribution, the distribution $P(s)$ represents 
a sort of average measure 
and tells us the extent of variation of the number of regions over the attractor with a given 
averaged local density. We find that this variation is, in general, similar 
for all chaotic attractors. The number of regions with very small average density dominating and 
decreasing as a power law as the average density increases, with very few regions having very high 
density. However, the specific details and the 
rate at which this variation occurs is characteristic for a given attractor and determined by the 
power law index $\gamma$. 

It should be noted that, by construction, the WRN and the associated measures are related to the 
structure of the embedded attractor and not directly to its dynamics. We illustrate this in 
Fig.~\ref{f.5} using the distributions derived from two different types of attractors. The top 
panel shows the normalized strength distribution of the WRN from a strange nonchaotic attractor 
(SNCA) \cite {greb,kapi} generated from a quasiperiodically forced pendulum given by:
\begin{equation}
{{d^2x} \over {dt^2}} + a{{dx} \over {dt}} + b\sin x  =  d +c(\cos \omega t + \cos \Omega t)
\label{eq:8}
\end{equation} 
with $a = 3.0$, $b = 1.0$, $c = 1.1$, $d = 1.33$, $\omega = (3 - \sqrt 5)/4$ and 
$\Omega = (1 + \sqrt 5)/4$. The distribution is similar to that from a chaotic attractor with 
power law scaling and exponential cut off. The value of $\gamma$ in this case is found to be 
$0.36 \pm 0.06$.  In other words, the power law scaling is actually a 
consequence of the fractal structure of the attractor and nothing to do with whether the 
system is chaotic or not. The connection between the scale free character of the distribution 
and the fractal structure of the attractor is a matter that requires a more detailed investigation. 
In the bottom panel, we show the typical strength distribution of the WRN from a periodic or 
quasiperiodic attractor. In this case, the strength is distributed over a small range and the 
number of nodes having strength within this range approximately remains constant. As $N$ 
increases, the range is found to shrink. 

In order to understand how noise affects the value of the exponent $\gamma$, we generated time 
series by adding different percentages of white noise to the Lorenz data. By computing the 
strength distribution, the $\gamma$ values are determned in each case and the results are 
shown in Fig.~\ref{f.6}. Though any clear scaling behavior cannot be deduced, it can be seen that 
the value of $\gamma$ tends to zero as the percentage of noise reaches $25\%$ and the distribution 
becomes exponential. 

\begin{figure}
\includegraphics[width=0.90\columnwidth]{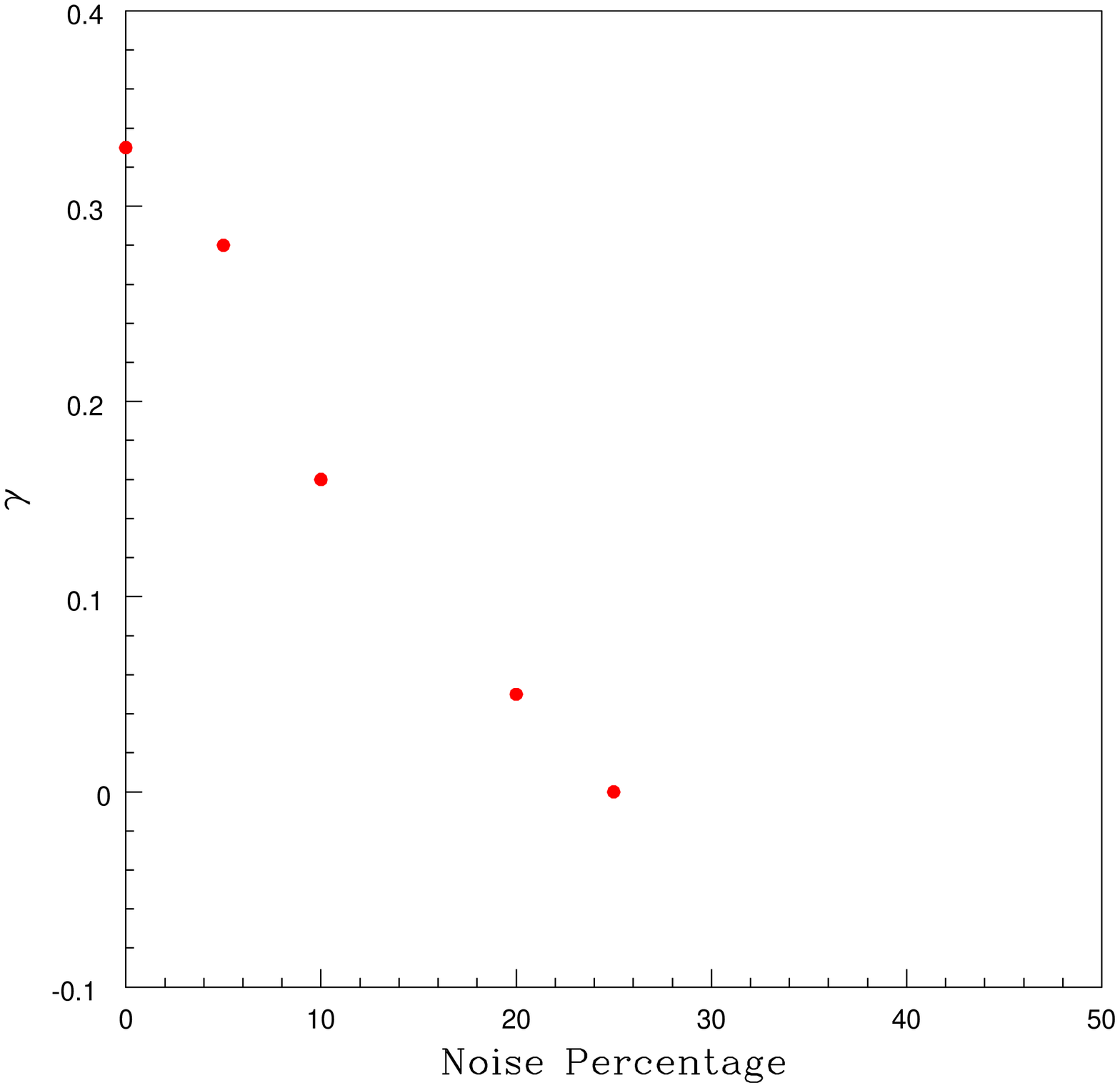}%
\caption{\label{f.6}(Color online)The decrease in the power law exponent $\gamma$ of the normalized 
strength distribution of WRN from Lorenz data with increasing percentage of white noise. }
\label{f.6}
\end{figure}

\subsection{Weighted clustering coefficient}
We now compute another primary network measure, the global clustering coefficient (CC) 
\cite {ops} for WRN. To compute CC for the unweighted complex network, we first define the local 
clustering coefficient of a node $\imath$ as 
\begin{equation}
C_i = {{2f_i} \over {k_i (k_i - 1)}}
\label{eq:9}
\end{equation}
where $k_i$ is the degree of the node and $f_i$ are the number of basic nontrivial motifs (triangles) 
attached to the node \cite {ops}. The value of $C_i$ measures how many of the nodes connected to the 
node $\imath$ are also mutually interconnected and its value is normalized in the range from 
$0$ to $1$. By averaging $C_i$ for all the nodes over the entire network, we get the global CC of 
the network. 

To generalize this, the weighted local CC of a node $\imath$, denoted by $C_i^w$, is first determined  
following Onnela et al. \cite {onn}. For this, we replace the number of triangles in Eq.~\ref{eq:9} 
with the sum of triangle intensities as 
\begin{equation}
C_i^w = {{{2} \over {k_i(k_i - 1)}} \sum_{j,l}({w_{ij}^n}{w_{jl}^n}{w_{li}^n})^{1/3}}
\label{eq:10}
\end{equation}
where the weight factors of the links are scaled by the largest weight factor in the network:
\begin{equation}
{w_{ij}^n} = {{w_{ij}} \over {max(w_{ij})}}
\label{eq:11}
\end{equation}
This definition also fulfills the requirement that $C_i^w \rightarrow C_i$ as the weights become binary. 
The average global weighted CC of the network is now given by
\begin{equation}
CC^W = {{{1} \over {N}} \sum_{i=1}^N C_i^W}
\label{eq:12}
\end{equation}

\begin{figure}
\includegraphics[width=0.64\columnwidth]{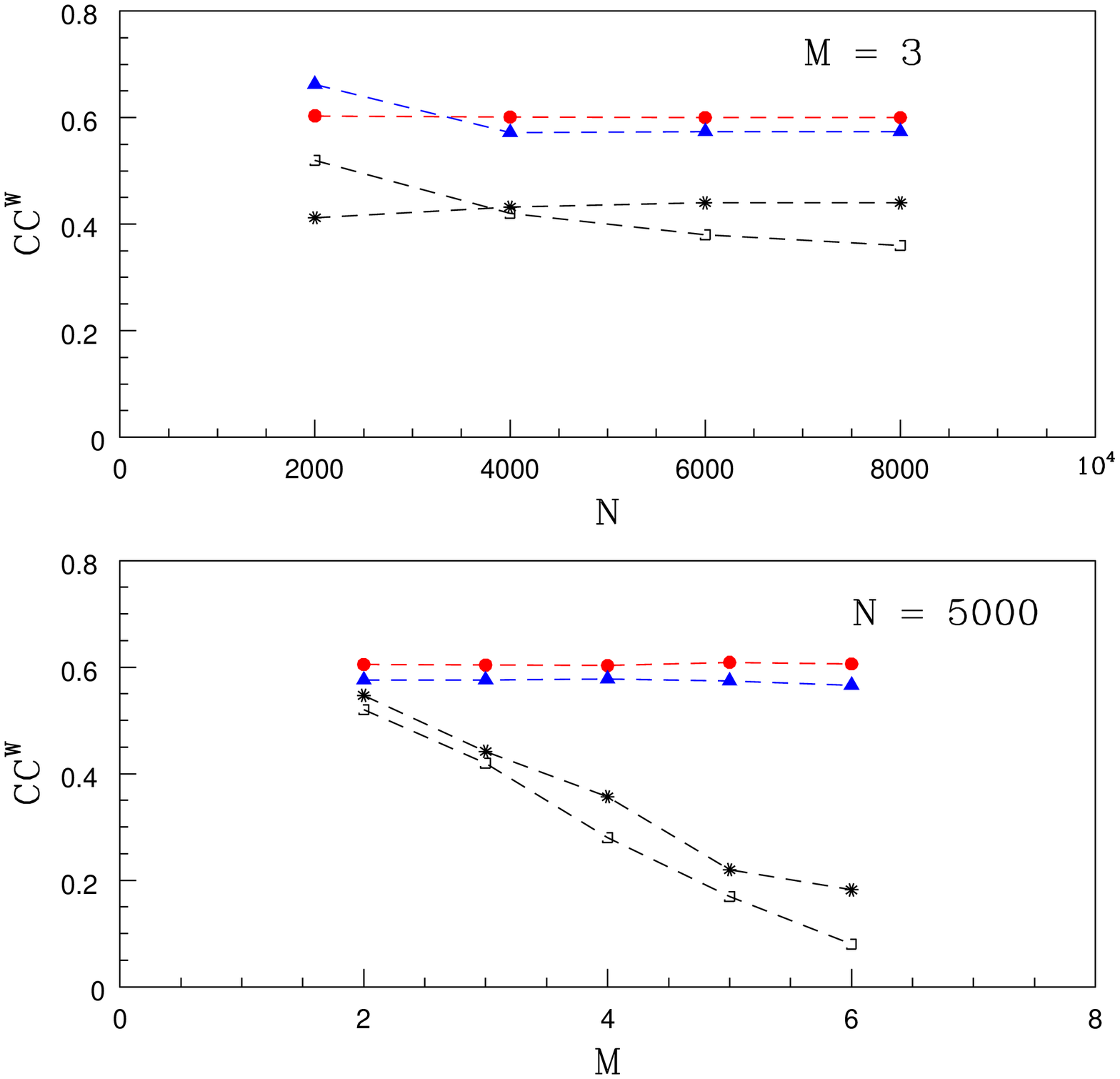}%
\caption{\label{f.7}(Color online)Top panel shows the variation of the clustering coefficient $(CC^W)$ 
of the weighted RN, as a function of $N$, constructed from Lorenz attractor (red solid circles), 
R\"ossler attractor (blue solid triangles), white noise (asterix) and $1/f$ noise (black open square). 
The value of $M$ is fixed at $3$. Bottom panel shows the same, but as a function of $M$ with $N$ 
fixed at $5000$.}
\label{f.7}
\end{figure}

We now compute the $CC^W$ for WRN from several standard chaotic attractors and random data by 
changing $N$ and $M$. The results are shown in Fig.~\ref{f.7} for the Lorenz and the R\"ossler data 
as well as the white and the $1/f$ noise. The top panel shows the results by fixing $M = 3$ and 
changing $N$ and while the bottom panel for fixed $N = 5000$ and changing $M$. The important 
result here is that, just like the strength distribution, $CC^W$ is also a characteristic measure 
for a given chaotic attractor independent of both $N$ and $M$. On the other hand, $CC^W$ for both 
white noise and $1/f$ noise show a decreasing trend as $M$ increases, though remains constant when 
$M$ is fixed and $N$ is increased. This is because, for the random data, the trajectory tends to 
fill the available dimension even as $M \rightarrow \infty$, unlike the case of a chaotic attractor. 
Thus, its clustering tends to decrease with $M$ for a fixed $N$. 

\subsection{Characteristic weight}
In this section, we focus on another important measure of any complex network, namely, the characteristic 
path length (CPL) and try to generalize this measure for the WRN that we consider here. We first briefly 
review the basic ideas for the unweighted case. The CPL is a measure of the global connectivity of a 
network and is defined through the shortest path length $l_s$ between any pair of nodes $(\imath,\jmath)$ 
in the network. Here $l_s$ represents the minimum number of nodes to be covered to reach from a reference 
node $\imath$ to any other node $\jmath$ in the network. To calculate CPL, we first compute $l_s$ for all 
the nodes $\jmath$ for a given $\imath$ and the average is found. This is repeated by changing $\imath$ 
for all the nodes in the network and the global average is found:
\begin{equation}
CPL = {{{1} \over {N}} \sum_{i=1}^N ({{1} \over {N-1}} \sum_{i \neq j=1}^{N-1} l_s)}
\label{eq:13}
\end{equation}

To generalise this for the weighted network, one should first note that the \emph {shortest path} has to 
be replaced by \emph {the path with the maximum weight factor}, since the connectivity increases with 
weight factor for a link. For example, suppose we consider the shortest path from a reference node 
$\imath$ to any other node $\jmath$, where $\imath$ and $\jmath$ are not directly connected. There will 
be different paths to reach $\jmath$ from $\imath$. The effective weight factor $w_{ij}$ for different 
paths will be different. We have to choose the path with the maximum of $w_{ij}$. To calculate the 
effective weight factor between nodes $\imath$ and $\jmath$, we use  
\emph {the geometric mean of the weight factors of all the links intermediate between $\imath$ and $\jmath$ for the path} \cite {ops}. This is repeated for all possible paths between $\imath$ and $\jmath$ 
and the highest effective weight factor is chosen as the characteristic weight between the nodes 
$\imath$ and $\jmath$ denoted as $w_{ij}^f$. 

Suppose  there are two possible 
paths between $1$ and $2$, namely, $1 \rightarrow 3 \rightarrow 2$ and 
$1 \rightarrow 4 \rightarrow 5 \rightarrow 2$. The effective weight factor for the former is given by 
\begin{equation}
{{1} \over {w_{12}^{(1)}}} = {{1} \over {w_{13}}} + {{1} \over {w_{23}}}
\label{eq:14}
\end{equation}
Or,    
\begin{equation}
{{w_{12}^{(1)}}} = {{w_{13} w_{23}} \over {w_{13} + w_{23}}}
\label{eq:15}
\end{equation}
Similarly, for the latter, the effective weight is
\begin{equation}
{{w_{12}^{(2)}}} = {{w_{14} w_{45} w_{52}} \over {w_{14} w_{45} + w_{45} w_{52} + w_{14} w_{52}}}
\label{eq:16}
\end{equation}

\begin{figure}
\includegraphics[width=0.90\columnwidth]{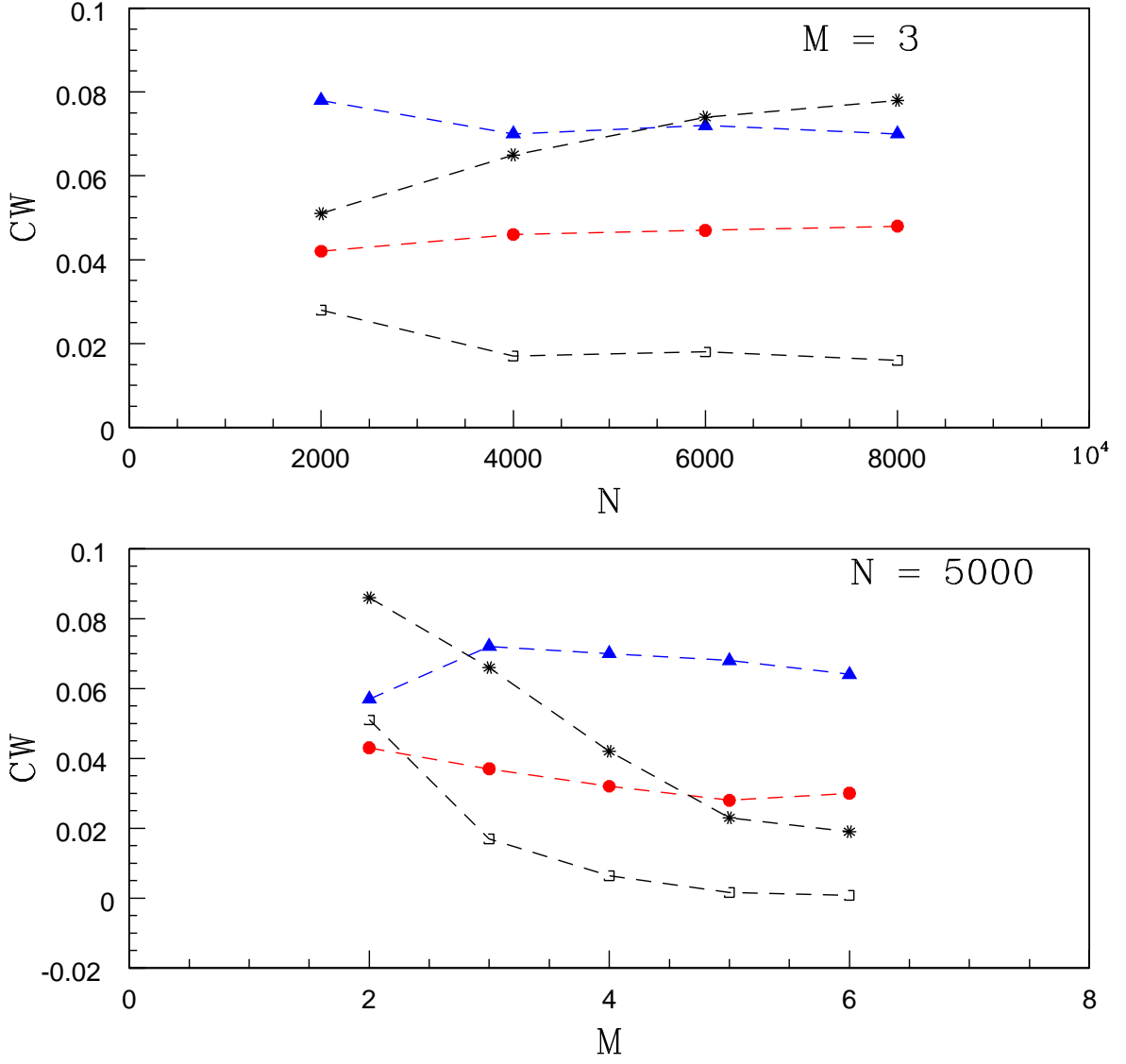}%
\caption{\label{f.8}(Color online)Variation of the characteristic weight (CW) of the weighted RN as a 
function of $N$ (top panel) for fixed $M$ and vice versa (bottom panel). Systems and symbols are the 
same as in Fig.~\ref{f.7}}
\label{f.8}
\end{figure}

The largest effective weight $w_{ij}^f$ is averaged by changing $\jmath$ for a given $\imath$ and 
for the whole network by changing $\imath$ from $1$ to $N$. We call it the 
\emph {characteristic weight} of the WRN denoted by $CW$:
\begin{equation}
CW = {{{1} \over {N}} \sum_{i=1}^N ({{1} \over {N-1}} \sum_{i \neq j=1}^{N-1} w_{ij}^f)}
\label{eq:17}
\end{equation}
It gives a measure for the global connectivity of the weighted network analogous to the CPL. 
However, a crucial difference between CPL and $CW$ should be noted. When the connections are 
binary, a higher value of CPL indicates that more number of nodes are to be covered, on the 
average, to reach from node $\imath$ to $\jmath$. In other words, the global connectivity depends  
inversely on CPL, which is always $> 1$. On the other hand, the weight factor $w_{ij}$ for a link is 
defined in such a way that a higher value of $CW$ (which is always $< 1$) indicates a better global 
connectivity for the whole network.

This measure is now computed for the WRN from various chaotic time series and noise. The results are 
shown in Fig.~\ref{f.8} for Lorenz, R\"ossler, white noise and $1/f$ noise. It is evident that 
the behavior of $CW$ is similar to that of $CC^W$. While it is a characteristic measure for a 
chaotic attractor, the same is not true for white noise and $1/f$ noise. For both, the global 
connectivity of the WRN decreases systematically as $M$ is increased. This also implies that the WRN 
provides an effective tool to distinguish  a chaotic time series from white noise and 
$1/f$ colored noise. This is shown in Fig.~\ref{f.9} where we present a combined plot of $CW$ and 
$CC^W$ for several standard chaotic attractors along with that for white and $1/f$ noise. The results 
indicate that WRN measures can be used effectively for the analysis of real world data.
  
\begin{figure}
\includegraphics[width=0.90\columnwidth]{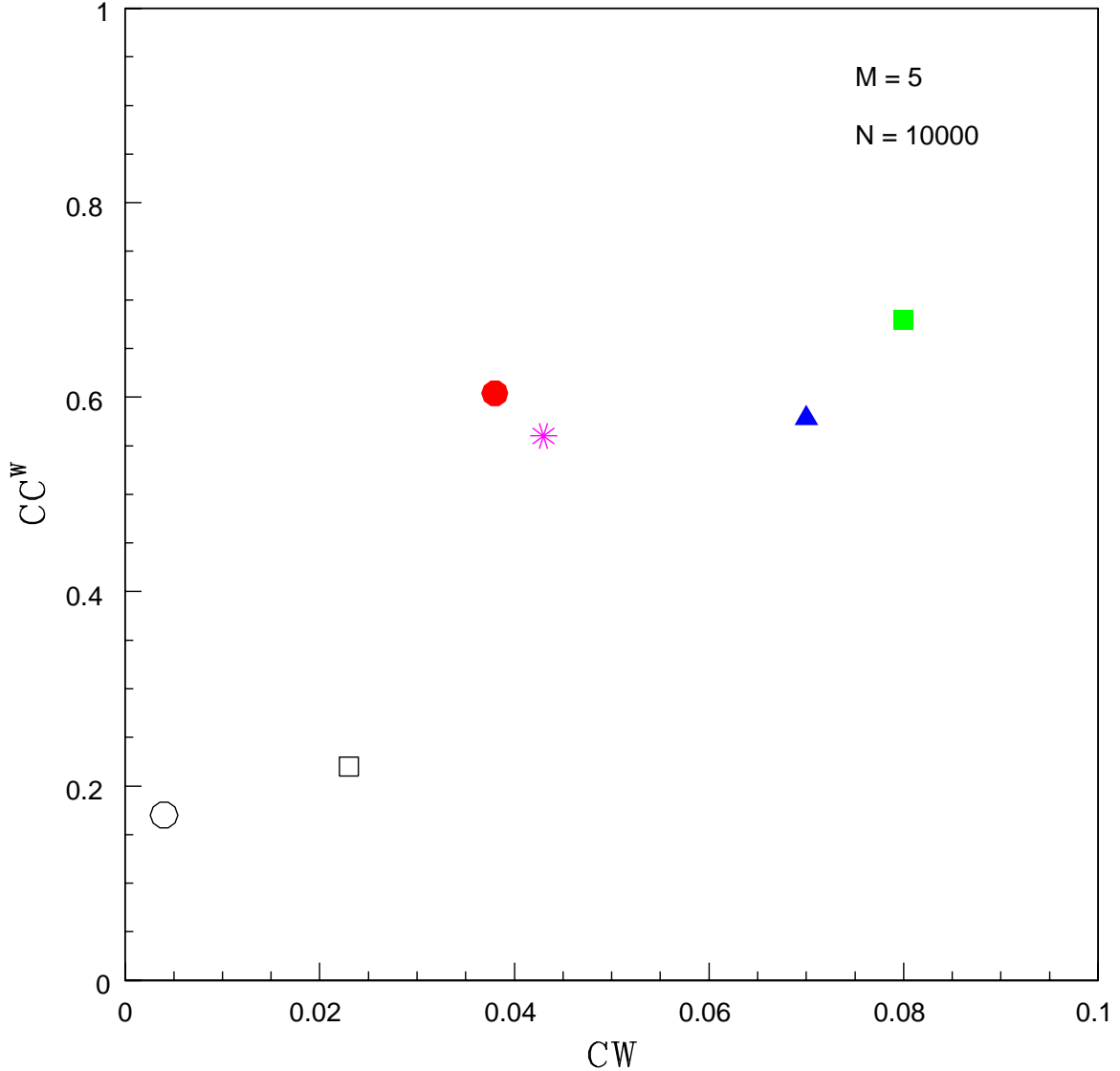}%
\caption{\label{f.9}(Color online)A combined plot of two WRN measures to distinguish chaos from white 
noise and $1/f$ noise. For chaotic systems, the average value from 10 different time series changing 
the initial conditions is plotted. For noise, the average value from 10 different simulations is taken. 
The values for Lorenz attractor (red solid circle), R\"ossler attractor (blue solid triangles), 
Henon attractor (green solid square), Ueda attractor (asterix), white noise (open square) and 
$1/f$ noise (open circle) are shown. The parameters used are $M = 5$ and $N = 5000$.}
\label{f.9}
\end{figure}

\section{\label{sec:level1}DISCUSSION AND CONCLUSION}
The analysis of time series data by converting into recurrence networks is an active area of research with applications over several domains. The recurrence networks considered so far in the literature are unweighted with binary connections. The basic objective of the work presented here is to propose a novel approach to construct a weighted recurrence network from a time series. We show how the constructed WRN and the associated measures have potential in nonlinear time series analysis, especially as discriminating measures to distinguish noisy from chaotic time series. This is achieved by generalizing three primary network measures so as to make them suitable for WRN and apply them to analyze time series from standard chaotic attractors, white noise and $1/f$ colored noise.    

We do not claim that the method proposed here is the only one to construct WRN from a time series. It may be possible to assign weight factors to links based on other strategies. Here we have used a general criterion for constructing the WRN, which in turn, provides some useful information regarding the structure of chaotic attractors.

We generalize three prominent network measures so as to make them suitable for WRN and apply them to 
analyze time series from standard chaotic attractors, white noise and $1/f$ colored noise. It is specifically shown that the node strength distribution of WRN from chaotic attractors follows a power law with exponential tail while that from white and $1/f$ noise is purely exponential. As the percentage of noise in the data increases, the power law part in the distribution depletes systematically and the distribution tends to exponential  with $\gamma \rightarrow 0$. 

Moreover, the weighted clustering coefficient ($CC^W$) and the characteristic weight ($CW$) of WRN from a chaotic attractor are invariants independent of 
all parameters involved in the construction of the network. On the other hand, both these measures systematically gets reduced with embedding dimension $M$ for white and $1/f$ noise, which implies that the local clustering and the global connectivity keeps on reducing with $M$ for noise. We foresee how the measures proposed here therefore can become important and powerful as characteristic measures for the analysis of real world data.  

An important aspect of the method is that it is possible to identify the WRNs from different chaotic attractors  as a single class with qualitatively similar strength distribution having power law scaling and exponential tail. Such a classification was not possible so far with the existing methods of undirected RNs. Moreover the power law index $\gamma$ is 
a manifestation of the fractal structure of the reconstructed attractor and can serve as a single unique index characterizing its geometric complexity. 

Added to this is the advantage that accurate results can be obtained with much less size of the time series or few observational data, which makes it useful in many contexts where lack of large data sets is a serious constraint in arriving at non-erroneous conclusions. Also, the method is general and can be adopted to generate weighted networks in other contexts like airport data, internet etc. with similar characterizations. We hope the methods and measures presented here will open up a new window in the pursuit of complexity in real world systems with immediate consequences in applied areas like climate data analysis \cite {boe1,boe2} as well as in quantifying structural complexity of attractors through network based approach.

\begin{acknowledgements}
RJ and KPH  acknowledges the computing facilities in IUCAA, Pune. 
\end{acknowledgements}

\end{document}